\documentclass[preprint,12pt]{elsarticle}

\usepackage{amssymb}
\usepackage{amsmath}
\usepackage{booktabs}
\usepackage{float}
\usepackage{adjustbox}
\usepackage{xcolor}
\usepackage{soul}

\journal{Journal of Power Scources}

\begin{document}

\begin{frontmatter}

%% Title, authors and addresses

%% use the tnoteref command within \title for footnotes;
%% use the tnotetext command for theassociated footnote;
%% use the fnref command within \author or \address for footnotes;
%% use the fntext command for theassociated footnote;
%% use the corref command within \author for corresponding author footnotes;
%% use the cortext command for theassociated footnote;
%% use the ead command for the email address,
%% and the form \ead[url] for the home page:
%% \title{Title\tnoteref{label1}}
%% \tnotetext[label1]{}
%% \author{Name\corref{cor1}\fnref{label2}}
%% \ead{email address}
%% \ead[url]{home page}
%% \fntext[label2]{}
%% \cortext[cor1]{}
%% \affiliation{organization={},
%%             addressline={},
%%             city={},
%%             postcode={},
%%             state={},
%%             country={}}
%% \fntext[label3]{}

\title{Comparative Performance Analysis of Numerical Discretization Methods for Electrochemical Model of Lithium-ion Batteries}

%% use optional labels to link authors explicitly to addresses:
%% \author[label1,label2]{}
%% \affiliation[label1]{organization={},
%%             addressline={},
%%             city={},
%%             postcode={},
%%             state={},
%%             country={}}
%%
%% \affiliation[label2]{organization={},
%%             addressline={},
%%             city={},
%%             postcode={},
%%             state={},
%%             country={}}

\author[inst1,inst2]{Feng Guo \corref{cor1}}
\author[inst1,inst2]{Luis D. Couto}

\affiliation[inst1]{organization={VITO},%Department and Organization
            addressline={Boeretang 200}, 
            city={Mol},
            postcode={2400}, 
            country={Belgium}}

\affiliation[inst2]{organization={EnergyVille},%Department and Organization
            addressline={Thor Park 8310}, 
            city={Genk},
            postcode={3600}, 
            country={Belgium}}

\cortext[cor1]{Corresponding author: feng.guo@vito.be (Feng Guo)}

\begin{abstract}
%% Text of abstract
This study evaluates numerical discretization methods for the Single Particle Model (SPM) used in electrochemical modeling. The methods include the Finite Difference Method (FDM), spectral methods, Padé approximation, and parabolic approximation. Evaluation criteria are accuracy, execution time, and memory usage, aiming to guide method selection for electrochemical models. Under constant current conditions, the FDM explicit Euler and Runge-Kutta methods show significant errors, while the FDM implicit Euler method improves accuracy with more nodes. The spectral method achieves the best accuracy and convergence with as few as five nodes. The Padé approximation exhibits increasing errors with higher current, and the parabolic approximation shows higher errors than the converged spectral and FDM implicit Euler methods. Under dynamic conditions, frequency domain analysis indicates that the FDM, spectral, and Padé approximation methods improve high-frequency response by increasing node count or method order. In terms of execution time, the parabolic method is fastest, followed by the Padé approximation. The spectral method is faster than FDM, while the FDM implicit Euler method is the slowest. Memory usage is lowest for the parabolic and Padé methods, moderate for FDM, and highest for the spectral method. These findings provide practical guidance for selecting discretization methods under different operating scenarios.

\end{abstract}

\begin{keyword}
%% keywords here, in the form: keyword \sep keyword
Lithium-ion batteries \sep Electrochemical model \sep Numerical Discretization \sep FDM \sep Spectral method \sep Padé approximation \sep Parabolic approximation

\end{keyword}

\end{frontmatter}

%% \linenumbers

%% main text
\section{Introduction}
\label{sec:intro}
With the widespread application of lithium-ion batteries, modeling, state estimation and control of these batteries have become increasingly important to ensure their performance and safety \cite{jiang2024advances,waseem2023battery,guo2024systematic}. The most commonly used battery model is the equivalent circuit model, due to its simple structure and ease of application \cite{li2021model}. However, as the demands for battery performance and safety increase, there is a growing need for a deeper understanding of the internal information of batteries. The equivalent circuit model, being merely a fit of the battery voltage, fails to provide extensive insights into the internal mechanisms of the battery, making it inadequate for higher demands. In contrast, electrochemical models based on the mechanisms of battery electrochemical reactions can accurately describe the internal reaction mechanisms of the battery \cite{ali2024comparison}. Therefore, research on electrochemical models is of great importance.

Electrochemical models typically refer to the Pseudo-Two-Dimensional (P2D) model and its variants. The P2D model was first proposed by Doyle, Fuller, and Newman, and is also known as the Doyle-Fuller-Newman (DFN) model \cite{fuller1994simulation}. The P2D model describes the microscopic electrochemical phenomena within a battery and establishes the connection between these processes and the macroscopic battery voltage, current, etc. This model provides deeper insights into the internal workings of batteries. The calculations and solutions of the P2D model are overly complex, leading to the development of many simplified models based on P2D. The most representative of these is the Single Particle Model (SPM)\cite{haran1998determination}. The SPM simplifies the P2D model in the following ways. First, it replaces the behavior of each electrode with a single particle representing the entire electrode. Then, it ignores the dynamics in the electrolyte, assuming a constant electrolyte concentration throughout the cell thickness. These simplifications significantly reduce the computational complexity of the battery model while retaining key internal information about the battery. Because the SPM  ignores the electrochemical processes in the electrolyte, it performs poorly under high current conditions. Therefore, an improved model, the SPMe, which includes the electrolyte dynamics, has been proposed \cite{marquis2019asymptotic}. In the process of solving the P2D model, SPM, and SPMe, it is necessary to solve the partial differential equation (PDE) that describes the diffusion of lithium in the solid phase of the electrode. Because it is difficult to obtain an analytical solution for this PDE, numerical discretization methods are typically used. The choice of method for solving this PDE directly affects the computational efficiency and accuracy of the electrochemical model. The common numerical spatial discretization methods used in electrochemical model can be divided into four types: finite methods, spectral methods, Padé approximation methods and parabolic approximation methods.

Finite methods generally include Finite Difference Method (FDM) \cite{di2010lithium,lotfi2020switched,xiong2018electrochemical,plante2022multiple,wang2023lithium}, Finite Volume Method (FVM) \cite{sturm2018state,li2023physics}, and Finite Element Method (FEM) \cite{tian2023aging,yeregui2023state}. Among these, FDM is typically used as a benchmark method. Many researchers have used FDM to solve electrochemical models because it is straightforward and allows for different numbers of nodes to balance computational speed and accuracy. When using FDM, it also requires a method to discretize the resulting ordinary differential equation in time, commonly using the explicit Euler method and the implicit Euler method. The implicit Euler method is more stable but requires more computation time, while the explicit Euler method is faster but less stable with a large number of nodes. In the field of solving electrochemical models, the Runge-Kutta method is also used to solve FDM. Xiong et al. proposed a Runge-Kutta method to solve a 10-node FDM model \cite{xiong2018electrochemical}. The Runge-Kutta method, similar to explicit methods, has the advantage of faster computation and is frequently used for numerical solutions of FDM. However, like explicit methods, the Runge-Kutta method can become unstable when solving with a large number of nodes.

Spectral methods transform the PDE into algebraic equations by expressing the solution as a linear combination of a set of global basis functions. Compared to finite methods, spectral methods result in a smaller system of equations for the same level of accuracy \cite{trefethen2000spectral}, which implies that spectral methods offer faster computation speeds for achieving the same accuracy. Spectral methods have also been employed in solving electrochemical models due to their advantageous properties, including stability and rapid computation \cite{bizeray2015lithium,wang2023system}. However, despite these benefits, spectral methods are more complex to implement compared to finite methods.

The Padé approximation method is based on representing a function as the ratio of two polynomials in frequency domain (i.e. transfer function), rather than using traditional polynomial approximations. This method has significant advantages in dealing with functions that exhibit complex behavior. It is often used for solving PDEs and has been widely applied in electrochemical models \cite{forman2010reduction,hosseininasab2023state}. Typically, in solving electrochemical models, the order of the Padé approximants used ranges from 2 to 5 \cite{forman2010reduction,hosseininasab2023state}.

Parabolic approximation method is also widely used for solving electrochemical models \cite{wang2023lithium, fang2023performance}. The diffusion process of lithium concentration within particles can be considered to follow a parabolic distribution\cite{subramanian2005efficient}. These methods are straightforward and can significantly simplify the solution process of lithium solid-phase diffusion in electrodes. However, unlike the FDM, the spectral method and the Padé approximation method, the parabolic approximation method does not allow for enhanced accuracy by increasing the number of nodes.

The specific choice of method for solving PDEs in electrochemical models directly affects the model's accuracy, execution speed, and memory usage. The requirements vary depending on the application. For example, in aging simulations, execution speed is the most critical factor. In battery management systems, due to hardware limitations, both execution speed and memory usage are of primary concern. If the aim is to study the battery's performance under extreme conditions, model accuracy becomes the top priority. Therefore, a comprehensive comparison of numerical discretization methods in terms of accuracy, execution speed, and memory usage is crucial for the application of electrochemical models.

Forman et al. compared the accuracy and computational time of the FDM using three different node selections and the Padé approximation method using three different orders \cite{forman2010reduction}. Their results indicated that the Padé approximation method is faster than FDM in terms of computational speed, but they did not provide numerical comparisons regarding accuracy. The limited number of nodes selected for FDM makes it challenging to conduct a comprehensive comparison between the two methods. Romero-Becerril et al. compared the accuracy of FDM, FVM, and differential quadrature method using the SPM model \cite{romero2011comparison}. However, the differential quadrature method is not commonly utilized for solving electrochemical models; researchers frequently opt for similar spectral methods, which offer superior convergence rates. Xu et al. compared the accuracy of the FDM and the FVM using the SPM \cite{xu2023comparative}. The study by Ali et al. compared the accuracy and computational speed of three models: P2D, SPMe, and SPM. Additionally, they compared two numerical discretization methods: parabolic approximation and Padé approximation\cite{ali2024comparison}. However, none of the above studies provide a comprehensive comparison of the four commonly used numerical discretization methods for solving electrochemical models, particularly in terms of accuracy, execution speed, and memory usage. Such a systematic comparison is crucial, as it can offer practical guidance for selecting appropriate discretization methods in different application scenarios. By understanding the trade-offs among these key performance metrics, researchers and engineers can make informed decisions to balance accuracy and computational efficiency in various electrochemical modeling tasks. 

Therefore, this study conducts a comparison of the FDM method, spectral method, the Padé approximation method, and parabolic approximation based on the SPM across accuracy, execution speed, and memory usage. Additionally, the study analyzes the impact of current magnitude and frequency domain analysis on the performance of these methods. This comprehensive evaluation offers valuable references for future researchers, enabling them to choose the most suitable numerical discretization method for solving electrochemical models based on their specific application requirements.

\section{SPM}
\label{sec:SPM}

This study uses the SPM to compare different numerical discretization methods. These numerical discretization methods are used to solve the diffusion of lithium in the solid phase of the electrode ( see Eq. \eqref{eq:1}). Eq. \eqref{eq:1} is a core component shared by all mainstream electrochemical models, including SPM, SPMe, and P2D. Although the models differ in other structural aspects, their treatment of this specific diffusion process is fundamentally similar. Therefore, when the focus is solely on evaluating the performance of discretization methods, the choice of model does not affect the results. Using the simpler SPM allows us to isolate the impact of the discretization methods without interference from model-specific complexities. The specific calculation formulas are as follows:
\begin{equation}
\frac{\partial c_{s,i}}{\partial t}(r,t) = \frac{D_{s,i}}{R_{s,i}^2} \frac{\partial}{\partial r} \left( r^2 \frac{\partial c_{s,i}}{\partial r}(r,t) \right)\label{eq:1}
\end{equation}
where  $i \in \{p,n\} $ with $p$ and $n$ as subscripts for positive and negative electrode, respectively, $c_{s}$ is the concentration of lithium in the solid phase, $D_{s}$ is the solid phase diffusion coefficient, $R_{s}$ is the particle radius, $r$ is the radial position within the electrode particle, and $t$ is time.

The boundary conditions for PDE Eq. \eqref{eq:1} are given at the center of the electrode particle by:
\begin{equation}
\left. \frac{\partial c_{s,i}}{\partial r}(r,t) \right\vert_{r=0} = 0\label{eq:2}
\end{equation}
and at the particle's surface  by:
\begin{equation}
\left. \frac{\partial c_{s,i}}{\partial r}(r,t) \right\vert_{r=R_s} = \frac{-j_i (t)}{D_{s,i}}  \label{eq:3}
\end{equation}
where $j_i$ is the flux of lithium ions into the particle.

In this study, the SPM variations compared only differ in their discretization methods used to solve Eq. \eqref{eq:1}, while all other complementary equations that define the output voltage are equivalent. 
The terminal cell voltage $V_{{\rm bat}}$ is calculated as:
\begin{equation}
V_{{\rm bat}} (t) = OCP_p(\theta_p(t)) - OCP_n(\theta_n(t)) + \eta_{p}(\theta_p(t), I(t)) - \eta_{n}(\theta_n(t), I(t)) - R_0 I(t)  \label{eq:4}
\end{equation}
where $OCP(\theta)$ represents the open-circuit potential, 
$\eta(\theta,I)$ is the surface overpotential, 
$I$ is the applied current 
and $R_0$ is the internal resistance of the cell. 
The variable $\theta_i$ is defined as:
\begin{equation}
\theta_i(t) = \frac{c_{ss,i}(t)}{c_{{\rm max},i}} \label{eq:5}
\end{equation}
where $c_{ss}$ is the solid-phase lithium concentration at the surface of the spherical particle and $c_{{\rm max}}$ is the maximum solid-phase lithium concentration. 
The overpotential functions are given by
%The remaining formulas of the SPM  are as follows:
\begin{equation}
\eta_{i}(\theta_i(t), I(t)) = \frac{2RT}{F} \sinh^{-1} \left( \frac{\pm I(t)}{2a_i L_i j_{0,i}(\theta_i(t))} \right) \label{eq:6}
\end{equation}
where 
$R$ is the universal gas constant and 
$T$ is the temperature. 
and $F$ is Faraday's constant.
$a$ is the specific surface area, $L$ is the thickness of the electrode and $\varepsilon_{s}$ is the volume fraction of the active material in the electrode. 
The exchange current density $j_{0,i}(\theta)$ is given by
\begin{equation}
% j_{0,i} = r_{eef,i} \sqrt{c_e c_{ss,i} (c_{max,i} - c_{ss,i})} \label{eq:5}
j_{0,i}(\theta_i(t)) = r_{{\rm eef},i} c_{{\rm max},i} \sqrt{c_e \theta_i(t) (1 - \theta_i(t))} \label{eq:7}
\end{equation}
where $r_{{\rm eef}}$ is the electrode reaction rate and $c_e$ is the lithium-ion concentration in the electrolyte.

The intercalation current density \( j_{i}(t) \) is assumed constant with respect to \( x \) along the cell thickness. This assumption makes it possible to represent \( j_{i}(t) \) as
\begin{equation}
j_{i}(t) = \pm\frac{1}{Fa_iAL_i} I(t)
\label{eq:8}
\end{equation}
where $A$ is electrode surface area.
$I(t)$ is the current at time $t$.

\section{Numerical Discretization Methods}
\label{sec:sample1}

Based on the SPM model introduced in Section \ref{sec:SPM}, this section explains how the FDM, spectral method, the Padé approximation method, and parabolic approximation method are used to solve Eq. \eqref{eq:1}. We rewrite Eq. \eqref{eq:1} into the state-space equations as follows:
\begin{eqnarray}
\dot{x}_{i}(t) &\!\!\!\!=&\!\!\!\! A_{i} \; x_{i}(t) + B_{i} \; j_i(t) \label{eq:9} \\
y_{i}(t) &\!\!\!\!=&\!\!\!\! C_{i} \; x_{i}(t) + D_{i} \; j_i(t) \label{eq:10}
\end{eqnarray}

\subsection{FDM}

The FDM is generally considered the benchmark method for spatially discretizing and numerically solving a PDE and it is widely used in solving electrochemical models of lithium-ion batteries. The basic idea of FDM is to divide the computational domain into discrete grid points and then approximate the differential operations at these grid points using finite differences. For the solid-phase lithium concentration diffusion equation in Eq. \eqref{eq:1}, a central difference for the second-order derivative is typically used \cite{xiong2018electrochemical,wang2023lithium}. The specific equation  after applying FDM is as follows:

The state vector $x_{\rm FDM}$ is defined as
\begin{equation}
x_{{\rm FDM},i}(t) = [ c_{s,i,1}(t) \;\; c_{s,i,2}(t) \;\; \cdots \;\; c_{s,i,(n-1)}(t) ]^\top
\label{eq:11}
\end{equation}
The state matrices are given by

\begin{equation}
A_{{\rm FDM},i} = \frac{D_{s,i}}{\Delta r_i^2}
\begin{bmatrix}
-2 & 2 & 0 & \cdots & 0 & 0 & 0 \\
\displaystyle \frac{1}{2} & -2 & \displaystyle \frac{3}{2} & \cdots & 0 & 0 & 0 \\
0 & \displaystyle \frac{2}{3} & -2 & \cdots & 0 & 0 & 0 \\
\vdots & \vdots & \vdots & \ddots & \vdots & \vdots & \vdots \\
0 & 0 & 0 & \cdots & \displaystyle \frac{n-3}{n-2} & -2 & \displaystyle \frac{n-1}{n-2} \\
0 & 0 & 0 & \cdots & 0 & \displaystyle  \frac{n-2}{n-1} & \displaystyle  -\frac{n-2}{n-1}
\end{bmatrix}\!\! \label{eq:12}
\end{equation}

\begin{equation}
B_{{\rm FDM},i} =
\frac{1}{\Delta r_i}
\begin{bmatrix}
0 \\ 
0 \\ 
\vdots \\
0 \\
\displaystyle -\frac{n}{n-1}
\end{bmatrix}\!\!
\label{eq:13}
\end{equation}

The output vector $y_{{\rm FDM},i}$ is defined as:
\begin{equation}
y_{{\rm FDM},i}(t) = [ c_{ss,i}(t) \;\; \bar{c}_{s,i}(t) ]^\top
\label{eq:14}
\end{equation}

And the output matrices are given by:

\begin{equation}
C_{{\rm FDM},i} = 
\begin{bmatrix}
0 & 0 & \cdots & 1 \\
\displaystyle \frac{\Delta r_i^3}{R_{s,i}^3} 
    & \displaystyle \frac{(2 \Delta r_i)^3 - (\Delta r_i)^3}{R_{s,i}^3} 
    & \cdots 
    & \displaystyle \frac{(n \Delta r_i)^3 - ((n-1) \Delta r_i)^3}{R_{s,i}^3}
\end{bmatrix}\!\!
\label{eq:15}
\end{equation}

\begin{equation}
D_{{\rm FDM},i} =
\begin{bmatrix}
\displaystyle - \frac{\Delta r_i}{D_{s,i}} \\ 
0
\end{bmatrix}\!\!
\label{eq:16}
\end{equation}
where $n$ represents the number of nodes, and $\Delta r_i$ represents the distance between two adjacent nodes. It is calculated by dividing the particle radius $R_{s,i}$ by $n$. 

Eq. \eqref{eq:11}  describes the solid-phase lithium concentration at all internal nodes of the spherical particle, excluding the surface nodes. 
The particle surface concentration $c_{ss}$ needs to be determined using the surface boundary conditions Eq. \eqref{eq:3}. The average lithium concentration in the particle $\bar{c}_{s}$. They are $y_{{\rm FDM},i}$ and determined via Eq. \eqref{eq:15} and Eq.\eqref{eq:16}.

Through the above steps, we used the FDM to spatially discretize Eq. \eqref{eq:1}, obtaining a continuous-time model. The next step is to discretize the model in time. Currently, the time discretization methods used in the FDM for electrochemical models include the explicit Euler method, the implicit Euler method, and the Runge-Kutta method

\subsubsection{Explicit Euler Method}

In the explicit Euler method, the state at the next time step can be directly computed from the current time step. For a given generic state-space model of the type in Eq. \eqref{eq:9}, the corresponding formulation is as follows:
\begin{equation}
x_i(t + \Delta t) = x_i(t) + \Delta t \left( A_i x_i(t) + B j_i(t) \right)
\label{eq:17}
\end{equation}
\begin{equation}
y_i(t + \Delta t) = C_i x_i(t + \Delta t) + D j_i(t + \Delta t)
\label{eq:18}
\end{equation}
where \(\Delta t\) is the sampling time. The main computational steps of this method involve matrix and vector multiplications and additions, which have relatively low computational complexity and can typically be completed quickly in a single iteration.

\subsubsection{Implicit Euler Method}

In the implicit Euler method, the state at the next time step involves solving a linear system of equations. The formula is as follows:
\begin{equation}
x_{i}(t + \Delta t) = (I - \Delta t \, A_{i})^{-1} \left( x_{i}(t) + \Delta t \, B_{i} \, j_i(t + \Delta t) \right) \label{eq:19}
\end{equation}
\begin{equation}
y_{i}(t + \Delta t) = C_{i} \, x_{i}(t + \Delta t) + D_{i} \, j_i(t + \Delta t) \label{eq:20}
\end{equation}

\subsubsection{Runge-Kutta Method}

In the Runge-Kutta method, the state at the next time step is computed using a weighted average of slopes calculated at intermediate points. This study uses 
 a third-order Runge-Kutta method \cite{xiong2018electrochemical}. The discrete-time state-space equations using the third-order Runge-Kutta method are:

\begin{equation}
k_1 = A_{i} \, x_{i}(t) + B_{i} \, j_i(t) \label{eq:21}
\end{equation}

\begin{equation}
k_2 = A_{i} \left( x_{i}(t) + \frac{\Delta t}{2} k_1 \right) + B_{i} \, \left( j_i(t) + \frac{\Delta t}{2} j_i(t) \right) \label{eq:22}
\end{equation}

\begin{equation}
k_3 = A_{i} \left( x_{i}(t) - \Delta t k_1 + 2 \Delta t k_2 \right) + B_{i} \, \left( j_i(t) - \Delta t j_i(t) + 2 \Delta t j_i(t) \right) \label{eq:23}
\end{equation}

Update the state:
\begin{equation}
x_{i}(t + \Delta t) = x_{i}(t) + \frac{\Delta t}{6} \left( k_1 + 4k_2 + k_3 \right) \label{eq:24}
\end{equation}

Update the output:
\begin{equation}
y_{i}(t + \Delta t) = C_{i} \, x_{i}(t + \Delta t) + D_{i} \, j_i(t + \Delta t) \label{eq:25}
\end{equation}

\subsection{Spectral Method}

The spectral method is considered an efficient and rapidly converging approach for solving PDEs. It achieves this by expressing the function over the entire solution domain as a linear combination of a set of basis functions. These basis functions are typically orthogonal polynomials, with Chebyshev polynomials being the most commonly used  \cite{trefethen2000spectral}. The implementation and derivation of the spectral method can be referenced from previous studies \cite{bizeray2015lithium,wang2023system}. In this study, we referenced the MATLAB code for the spectral method from previous researchers \cite{bizeray_2016_212178} and rewrote it in Python to match the format of the previously described methods. 

Depending on the selected order for the spectral approximation, the coefficients inside the model matrices are different and thus an $n$-th order general state-space representation cannot be provided, since it has to be derived. Just as an example, the 2nd order spectral model is given by:

\begin{equation}
A_{{\rm SM},i} = 
\begin{bmatrix}
0 & 0  \vspace{0.15cm} \\
0 & \displaystyle -20\frac{D_{s,i}}{R_{s,i}^{2}}
\end{bmatrix}\!\!\;\;
\label{eq:26}
\end{equation}

\begin{equation}
B_{{\rm SM},i} =
\begin{bmatrix}
\displaystyle -3 \vspace{0.15cm} \\ 
\displaystyle 2.18
\end{bmatrix}\!\!
\label{eq:27}
\end{equation}

The output vector $y_{{\rm SM},i}$ is defined as

\begin{equation}
y_{{\rm SM},i}(t) = [ c_{ss,i}(t) \;\; \bar{c}_{s,i}(t) ]^\top
\label{eq:28}
\end{equation}

And the output matrices are given by

\begin{equation}
C_{{\rm SM},i} = 
\displaystyle \frac{1}{R_{s,i}}
\begin{bmatrix}
1 & -0.98 \vspace{0.15cm} \\
1 & 0
\end{bmatrix}\!\!
\label{eq:29}
\end{equation}

\begin{equation}
D_{{\rm SM},i} =
\displaystyle \frac{R_{s,i}}{D_{s,i}}
\begin{bmatrix}
\displaystyle -0.094 \vspace{0.15cm} \\ 
\displaystyle 0
\end{bmatrix}\!\!
\label{eq30}
\end{equation}

This study uses the Zero-Order Hold (ZOH) method for temporal discretization in the Spectral method, a commonly used technique for dynamic systems to ensure accurate time discretization. The ZOH method provides several advantages, including simplicity of implementation and the ability to maintain a constant input signal within each sampling interval, which is especially beneficial in digital control systems \cite{pechlivanidou2022zero}.

\subsection{Padé Approximation Method}

The first step to obtain a Padé approximation of the PDE Eq. \eqref{eq:1} is to take the Laplace transform of this equation to obtain a trascendental transfer function in the frequency domain. This function can then be truncated through Padé  approximation for a given model order. 
Detailed derivations can be found in previous research works \cite{marcicki2013design,moura2014adaptive,zhang2022beyond,li2022unlocking}. Earlier studies often used 2nd to 4th order Padé approximations, whereas in this study we considered from 2nd to 5th order Padé approximations for comparison sake.

A given generic model of $n$th order obtained from a Padé approximation can be written the following transfer function $G(s)$
\begin{equation}
    G(s) = \frac{P(s)}{Q(s)} = \frac{p_{m-1} s^{m-1} + p_{m-2} s^{m-2} + \cdots + p_1 s + p_0}{s (s^{n-1} + q_{n-1} s^{n-2} +\cdots + q_2 s^2 + q_1)}
    \label{eq:31}
\end{equation}
where $s$ is the Laplace variable and $p$ and $q$ are the numerator and denominator coefficients of the polynomial functions $P(s)$ and $Q(s)$, respectively. This model can then be transformed from the transfer function Eq. \eqref{eq:31} to a controllable canonical stats-space form given by:

\begin{equation}
A_{{\rm PdAM},i} = 
\begin{bmatrix}
0 & 1 & 0 & \cdots & 0 \\
0 & 0 & 1 & & 0 \\ 
\vdots & & & \ddots & \vdots \\
0 & 0 & 0 & \cdots & 1 \\
0 & -q_1 & -q_2 & \cdots & -q_{n-1}
\end{bmatrix}\!\!
\label{eq:32}
\end{equation}

\begin{equation}
B_{{\rm PdAM},i} =
-\frac{3}{R_{s,i}} 
\begin{bmatrix}
0 \\
0 \\
\vdots \\
0 \\
1
\end{bmatrix}\!\
\label{eq:33}
\end{equation}

The output vector $y_{{\rm PdAM},i}$ is defined as:
\begin{equation}
 y_{{\rm PdAM},i}(t) = [ \bar{c}_{s,i}(t) \;\; c_{ss,i}(t) ]^\top
\label{eq:34}
\end{equation}
 
and the output matrices are given by:

\begin{equation}
C_{{\rm PdAM},i} = 
\begin{bmatrix}
q_1 & q_2 & q_3 & \cdots & q_{n-1} & 1 \\
p_0 & p_1 & p_2 & \cdots & p_{n-2} & p_{n-1}
\end{bmatrix}\!\!
\label{eq:35}
\end{equation}

\begin{equation}
D_{{\rm PdAM},i} =
\begin{bmatrix}
0 \\ 
0
\end{bmatrix}\!\!
\label{eq:36}
\end{equation}

Depending on the selected order for the Padé approximation, the values of the coefficients will change accordingly and therefore a general state-space representation of order $n$ cannot be given but it has to be derived. Just as an example, the 2nd order Padé approximation model is given by involves the following matrices:

\begin{equation}
A_{{\rm PdAM},i} = 
\begin{bmatrix}
    0 &1 \\
    0 &\displaystyle -\frac{35 D_{s,i}}{R_{s,i}^2}
\end{bmatrix}\!\! \;\;
\label{eq:37}
\end{equation}

\begin{equation}
C_{{\rm PdAM},i}^\top =
\begin{bmatrix}
    \displaystyle \frac{35 D_{s,i}}{R_{s,i}^2} &\displaystyle \frac{35 D_{s,i}}{R_{s,i}^2} \vspace{0.1cm} \\
    1 &\displaystyle \frac{10}{3}
\end{bmatrix}\!\
\label{eq:38}
\end{equation}

The time discretization solutions for Padé approximation method also use the ZOH method.

\subsection{Parabolic Approximation Method}
The parabolic approximation method uses a parabola to approximate the process described in  Eq. \eqref{eq:1}, transforming it into an ordinary differential equation for solving \cite{subramanian2005efficient}. The calculation formulas for the parabolic approximation method are as follows \cite{tagade2016recursive,fan2020systematic}:

The state vector $x_{{\rm PrAM},i}$ is defined as:
\begin{equation}
x_{{\rm PrAM},i}(t) = [ \overline{c}_{s,i} \;\;  \overline{c}_{fs,i}]^\top
\label{eq:39}
\end{equation}
where $\overline{c}_{fs}$ is the average concentration flux of lithium in active material.
The state matrices are given by:

\begin{equation}
A_{{\rm PrAM},i} = 
\begin{bmatrix}
0 & 0  \vspace{0.15cm} \\
0 & \displaystyle -30\frac{D_{s,i}}{R_{s,i}^{2}}
\end{bmatrix}\!\!\;\; 
\label{eq:40}
\end{equation}

\begin{equation}
B_{{\rm PrAM},i} =
\begin{bmatrix}
\displaystyle -\frac{3}{R_{s,i}} \vspace{0.15cm} \\ 
\displaystyle -\frac{45}{2R_{s,i}^{2}}
\end{bmatrix}\!\!
\label{eq:41}
\end{equation}

The output vector $y_{{\rm PrAM},i}$ is defined as:
\begin{equation}
 y_{{\rm PrAM},i}(t) = [ c_{ss,i}(t) \;\; \bar{c}_{s,i}(t) ]^\top
\label{eq42}
\end{equation}

And the output matrices are given by:

\begin{equation}
C_{{\rm PrAM},i} = 
\begin{bmatrix}
1 & 0 \vspace{0.15cm} \\
1 & \displaystyle \frac{8R_{s,i}}{35}
\end{bmatrix}\!\!
\label{eq:43}
\end{equation}

\begin{equation}
D_{{\rm PrAM},i} =
\begin{bmatrix}
0 \vspace{0.15cm} \\ 
\displaystyle -\frac{R_{s,i}}{35D_{s,i}}
\end{bmatrix}\!\!
\label{eq:44}
\end{equation}

For the parabolic approximation method, this study also uses the ZOH method to discretize time.

\section{Results and Discussion} \label{sec:R&D}
In this section, we first present the implementation details of the different discretized models followed by the results of the battery model parameter identification. Then, we discuss the accuracy, computation time, and memory usage of different numerical discretization methods.

\subsection{Simulation Details}

All battery models and test codes in this study were implemented in Python, version 3.9.18. They were developed using a consistent program structure and identical data types to store the same variables. The sampling time is 1s for all the model. The execution time evaluation was conducted using the \texttt{time} module. The execution time calculation includes only the time taken by each method to run the corresponding model, excluding the time for loading test cases and data. The tests were executed on a computer with an Intel(R) Core(TM) i5-1145G7 processor, 16GB of memory, and running Windows 10 Enterprise. To account for variations in computation time during each run, the average execution time of 10 runs was used as the final computation time.

The error in battery models originates from the structural error and the parameter error of the battery model. Structural error refers to the unmodeled battery dynamics relative to the real battery. Parameter error refers to the inaccuracies in the identified parameters. It is challenging to completely eliminate these two types of errors in battery modeling and parameter identification. Therefore, they only need to be controlled within an acceptable range. Since this study focuses on the errors introduced by numerical discretization methods, this study uses a model with 200 nodes based on the implicit solution of the FDM as the benchmark model, referred to as \( \mathrm{FDM_{implicit}(200)} \). The FDM implicit method is one of the most widely used and recognized stable numerical schemes for solving diffusion-type PDEs, and is commonly adopted in electrochemical modeling due to its robustness and accuracy. In theory, increasing the number of discretization nodes improves the numerical accuracy of the solution. To ensure sufficient spatial resolution, we selected a relatively large number of nodes 200. 

Additionally, for methods like FDM, Spectral Method, and Padé Approximation Method, which allow the selection of different nodes or orders, accuracy is compared at various nodes or orders. It is important to note that using the Explicit Euler Method and Runge-Kutta Method, the FDM becomes unstable and fails when the number of nodes increase. In this study, the Explicit Euler Method fails when the number of nodes exceeds 14, and the Runge-Kutta Method fails when the number of nodes exceeds 15. Therefore, for these two methods, only the nodes for which the solutions are successfully obtained are compared.

\subsection{Model Parameter Identification}

The battery used in this study is a 2.9Ah commercial 18650 lithium-ion battery with an Nickel Manganese Cobalt (NMC) cathode and a graphite anode. The method used for battery model parameter identification is Particle Swarm Optimization (PSO), which is known for its stability and accuracy \cite{guo2024efficiency}. The model used for battery parameter estimation is \( \mathrm{FDM_{implicit}(200)} \). The battery test data used for parameter identification includes constent discharge data at  0.2C, 0.333C, 0.5C, 1C, and 3C rates. The parameter estimation ranges \cite{fan2020systematic} and the corresponding estimated values used in this study are presented in Table \ref{tab:1}.

\begin{table}[H]
\centering
\caption{Parameters Estimated in the Mode}
\resizebox{1.2\textwidth}{!}{
\begin{tabular}{cccc}
\hline
\textbf{Variable} & \textbf{Parameter Definition} & \textbf{Range} & \textbf{Estimated Value} \\ \hline
\(c_{n,0}\)  & Initial concentration in negative electrode [${\rm mol}/{\rm m}^3$] & [30000, 50000] & 34947.74 \\
\(c_{p,0}\) & Initial concentration in positive electrode [${\rm mol}/{\rm m}^3$] & [0, 1000] & 331.33 \\
\(r_{eef,n}\)  & Negative electrode reaction rate [-] & [$1\times10^{-8}$, $1\times10^{-5}$] & $3.12\times10^{-7}$ \\
\(r_{eef,p}\) & Positive electrode reaction rate [-] & [$1\times10^{-8}$, $1\times10^{-5}$] & $4.28\times10^{-6}$ \\
\(\varepsilon_{s,n}\) & Negative electrode active material volume fraction [-] & [0.35, 0.7] & 0.5 \\
\(\varepsilon_{s,p}\) & Positive electrode active material volume fraction [-] & [0.35, 0.7] & 0.43 \\
\(D_{s,n}\) & Negative electrode diffusivity [${\rm m}^2/{\rm s}$] & [$1\times10^{-15}$, $1\times10^{-10}$] & $1.34\times10^{-14}$ \\
\(D_{s,p}\) & Positive electrode diffusivity [${\rm m}^2/{\rm s}$] & [$1\times10^{-15}$, $1\times10^{-10}$] & $1.48\times10^{-14}$ \\
\(R_{0}\)  & Internal resistance [$\Omega$] & [0.001, 0.1] & 0.016 \\
\(c_{{\rm max},n}\) & Maximum concentration in negative electrode [${\rm mol}/{\rm m}^3$] & [35000, 50000] & 39139.31 \\
\(c_{{\rm max},p}\) & Maximum concentration in positive electrode [${\rm mol}/{\rm m}^3$] & [35000, 50000] & 46248.50 \\
\(L_n\) & Negative electrode thickness [\rm m]& [$1\times10^{-5}$, $1\times10^{-4}$] & $7.60\times10^{-5}$ \\
\(L_p\) & Positive electrode thickness [\rm m]& [$1\times10^{-5}$, $1\times10^{-4}$] & $6.94\times10^{-5}$ \\
\(R_{s,n}\) & Negative particle radius [\rm m]& [$1\times10^{-6}$, $1\times10^{-5}$] & $6\times10^{-6}$ \\
\(R_{s,p}\) & Positive particle radius [\rm m] & [$1\times10^{-6}$, $1\times10^{-5}$] & $4\times10^{-6}$ \\
\rm A & Electrode surface area [$\rm m^2$] & [0.006, 0.012] & 0.008 \\
\hline
\end{tabular}
}

\label{tab:1}
\end{table}

The RMSE obtained after the parameter identification process for each of the considered operating condition is listed in Table \ref{tab:2}. It can be seen that the average RMSE of the \( \mathrm{FDM_{implicit}(200)} \) model across different conditions is 32.37mV. In the following research, this  \( \mathrm{FDM_{implicit}(200)} \)  model is used as the benchmark model for comparing the accuracy of different models.

\begin{table}[H]
    \centering
    \caption{Model parameter identification error} % Sets the caption of the table to "Model Parameter Identification Error"
    \label{tab:2} % Sets a label for the table, which can be used for referencing it in the text
    \begin{tabular}{@{}lcccccc@{}} % Starts the tabular environment. @{} removes the extra space at the beginning and end of the table.
        \toprule % Adds a top horizontal line which is thicker and more spaced than the default line
        C rate & 0.2C  & 0.333C & 0.5C & 1C & 3C & Average \\ % Table header
        \midrule % Adds a middle horizontal line
        RMSE(mV) & 18.99 & 20.68 & 34.02 & 42.70 & 45.37 & 32.37 \\ % Table row with data
        \bottomrule % Adds a bottom horizontal line which is thicker and more spaced than the default line
    \end{tabular}
\end{table}

\subsection{Accuracy}
In this section, we first explain how the accuracy is evaluated and compared across methods. Then, we present the accuracy results of the all numerical discretization methods considered.

\subsubsection{Accuracy Evaluation}

\begin{figure}[H]
    \noindent
    \makebox[\linewidth]{\includegraphics[width=1.1\textwidth]{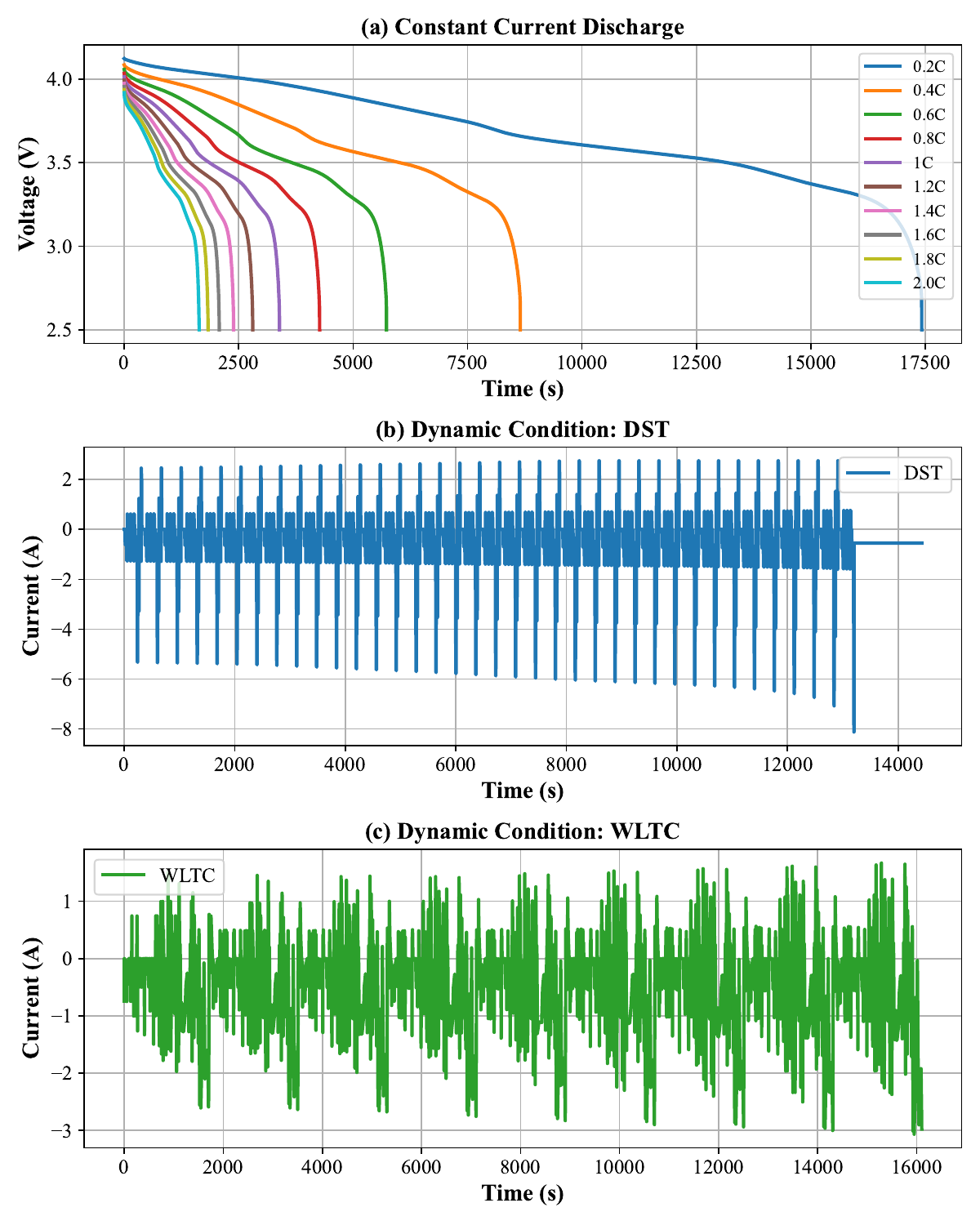}}
    \vspace{-0.8cm}
    \caption{Current input conditions used for comparison: (a) Constant current, (b) Dynamic Stress Test (DST), and (c) Worldwide harmonized Light vehicles Test Cycle (WLTC).}
    \label{fig:1}
\end{figure}

This study divides the accuracy comparison into two parts: accuracy under constant current  conditions and accuracy under dynamic conditions. Detailed information on the two type operating conditions can be found in Figure \ref{fig:1}. For the constant current conditions, we use \( \mathrm{FDM_{implicit}(200)} \) as the baseline model to generate voltage simulation data at 0.2C, 0.4C, 0.6C, 0.8C, 1C, 1.2C, 1.4C, 1.6C, 1.8C, and 2C (see Figure \ref{fig:1}(a)). We then compare the accuracy of different methods under these constant current conditions. The accuracy is evaluated using the mean absolute error (MAE), which is less sensitive to individual extreme errors than RMSE and better reflects the overall error. The specific calculation formula for MAE is as follows:

\begin{equation}
\mathrm{MAE} = \frac{1}{n_T} \sum_{i=1}^{n_T} |\mathrm{FDM_{implicit}(200)}_i - \hat{y}_i|
\label{eq:45}
\end{equation}
where:
\begin{itemize}
    \item $n_T$ is the number of samples,
    \item $\mathrm{FDM_{implicit}(200)}_i$ is the \( \mathrm{FDM_{implicit}(200)} \) voltage value for the $i$-th sample,
    \item $\hat{y}_i$ is the predicted voltage value for the $i$-th sample.
\end{itemize}

For dynamic conditions, we want to evaluate the performance of different methods under varying current dynamics, for which we considered Dynamic Stress Test (DST)  (see Figure \ref{fig:1}(b)) and Worldwide harmonized Light vehicles Test Cycle (WLTC) (see Figure \ref{fig:1}(c)). However, directly using the DST or WLTC the for assessment introduces the influence of different current magnitudes. To more accurately evaluate the impact of battery models under dynamic conditions, this study compares different models using frequency domain responses. This approach allows for a straightforward comparison of the losses of different models at specific frequencies.

\subsubsection{Constant Current Conditions}

\begin{figure}[H]
    \noindent
    \makebox[\linewidth]{\includegraphics[width=1.2\textwidth]{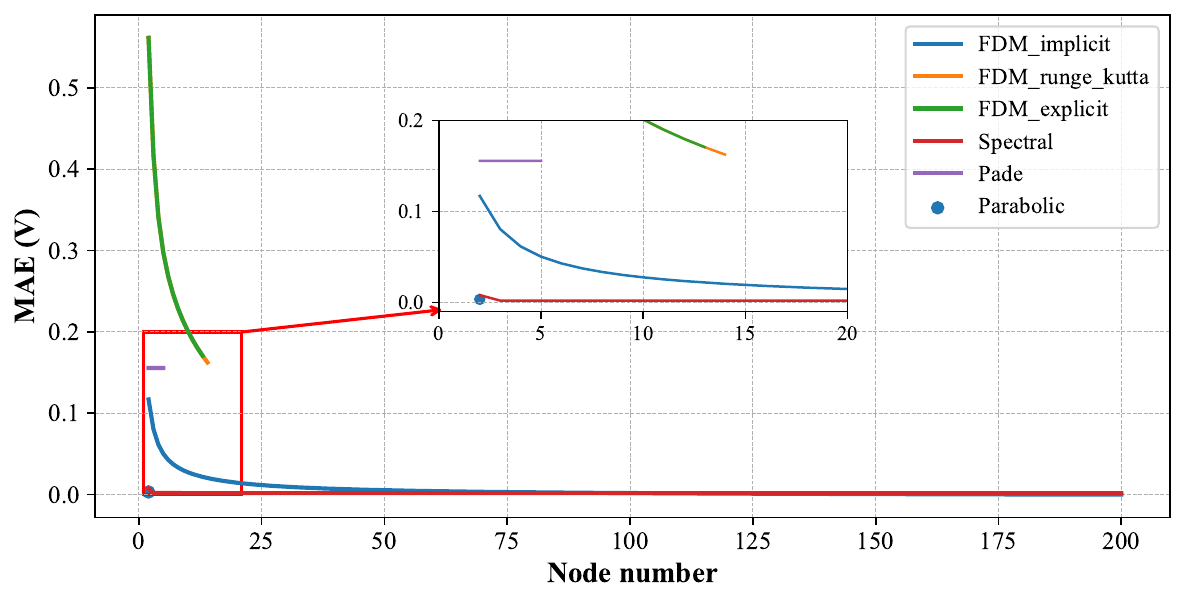}}
    \vspace{-0.8cm}
    \caption{Accuracy results with different nodes.} 
    \label{fig:2}
\end{figure}

Under constant current conditions, a comparison of the accuracy of different methods is illustrated in  Figure \ref{fig:2}. The horizontal axis represents the number of nodes, ranging from 2 to 200. The vertical axis indicates the average MAE across different current magnitudes. Specific error values for each method with the number of nodes ranging from 0 to 20 are detailed in Table \ref{tab:3}.

\begin{table}[ht]
    \centering
    \caption{Average MAE(mV) for different methods.}
    \begin{tabular}{ccccccc}
        \toprule
        Node & Implicit  & Runge Kutta  & Explicit & Spectral & Padé & Parabolic \\
        \midrule
        2  & 117.11 & 561.62 & 561.65 & 8.015 & 155.64 & 3.307 \\
        3  & 80.50 & 417.00 & 417.05 & 1.939 & 155.64 &  \\
        4  & 61.52 & 341.52 & 341.52 & 1.930 & 155.64 &  \\
        5  & 50.29 & 297.79 & 297.79 & 1.929 & 155.64 &  \\
        6  & 42.83 & 268.43 & 268.43 & 1.929 &  &  \\
        7  & 37.47 & 246.45 & 246.45 & 1.929 &  &  \\
        8  & 33.39 & 228.80 & 228.80 & 1.929 &  &  \\
        9  & 30.16 & 213.97 & 213.97 & 1.929 &  &  \\
        10 & 27.52 & 201.17 & 201.17 & 1.929 &  &  \\
        11 & 25.32 & 189.90 & 189.90 & 1.929 &  &  \\
        12 & 23.45 & 179.85 & 179.85 & 1.929 &  &  \\
        13 & 21.84 & 170.79 & 170.79 & 1.929 &  &  \\
        14 & 20.43 & 162.58 &  & 1.929 &  &  \\
        15 & 19.18 &  &  & 1.929 &  &  \\
        16 & 18.08 &  &  & 1.929 &  &  \\
        17 & 17.09 &  &  & 1.929 &  &  \\
        18 & 16.20 &  &  & 1.929 &  &  \\
        19 & 15.39 &  &  & 1.929 &  &  \\
        20 & 14.65 &  &  & 1.929 &  &  \\
        \bottomrule
    \end{tabular}
    \label{tab:3}
\end{table}

In the FDM and spectral method, the error decreases as the number of nodes increases. The errors of the explicit Euler method and the Runge-Kutta method are similar and both are greater than those of the implicit Euler Method. This study uses a sampling time of 1 second. Reducing the sampling time can further improve the accuracy of these two methods. The Padé approximation method does not exhibit a significant reduction in error with increasing order under constant current conditions.

From the comparison, it is evident that the spectral method converges much faster (with fewer nodes) than the FDM implicit Euler method, achieving very high accuracy with just 5 nodes. The accuracy of the parabolic approximation method is comparable to that of the FDM implicit Euler method with 72 nodes. Notably, the error of the spectral method with 5 nodes is smaller than that of the parabolic approximation method, and equivalent to the error of the FDM implicit Euler method with 100 nodes.

Based on these results, under constant current conditions, the FDM explicit Euler method and the Runge-Kutta method exhibit the largest errors and are less stable. Conversely, the spectral method shows faster convergence and achieves high accuracy with just 5 nodes. Increasing the order of the Padé approximation method does not effectively enhance its accuracy. The parabolic approximation method’s accuracy is lower than that of both the converged spectral method and the FDM implicit Euler method.

\begin{figure}[H]
    \noindent
    \makebox[\linewidth]{\includegraphics[width=1.1\textwidth]{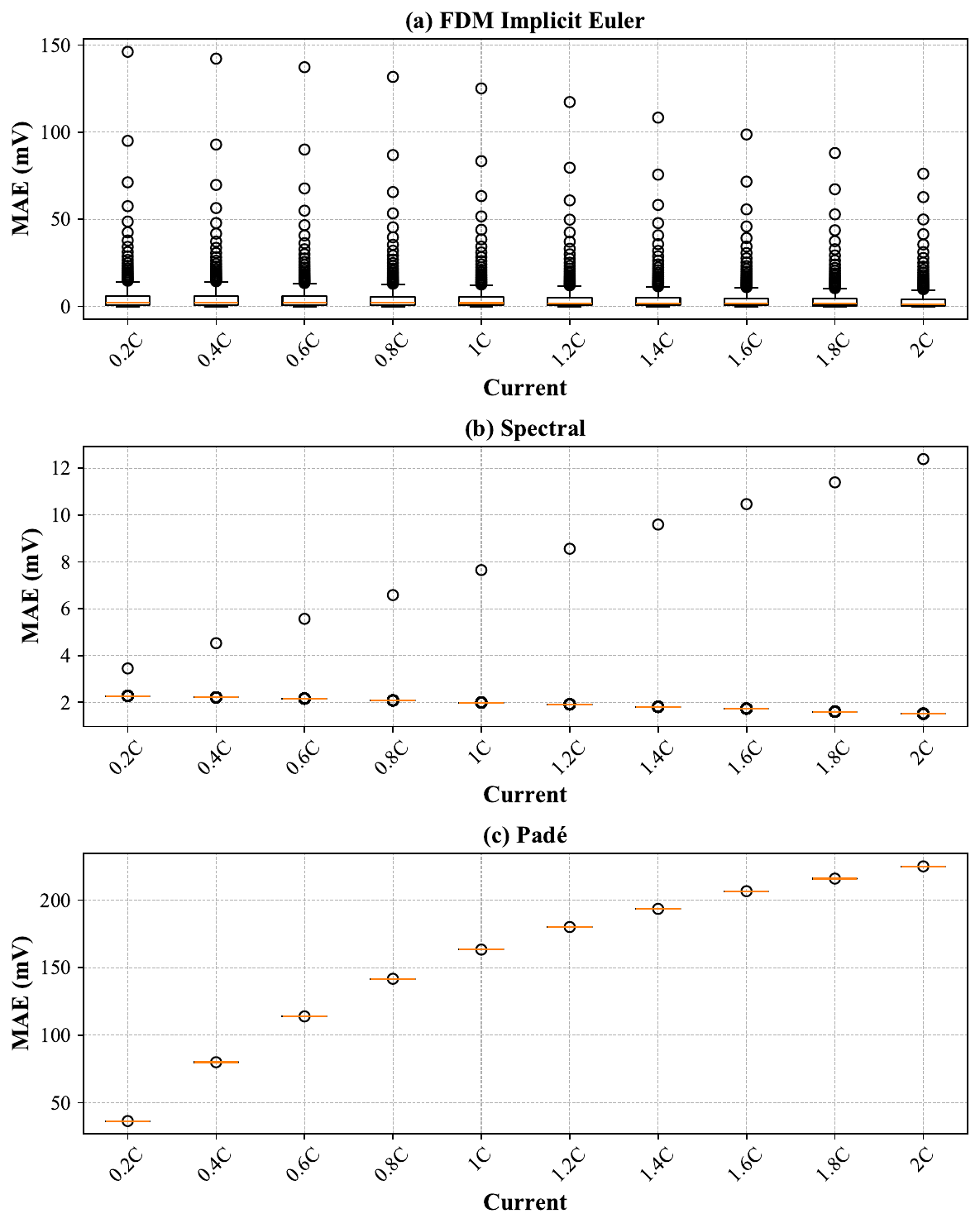}}
    \vspace{-0.8cm}
    \caption{MAE distribution across various conditions and node settings for (a) FDM with Implicit Euler, (b) Spectral Method, and (c) Padé Approximation.}
    \label{fig:3}
\end{figure}

Figure \ref{fig:3}(a) shows the box plot distribution of MAE for the FDM implicit Euler method under different currents and nodes. The horizontal axis represents different current magnitudes, arranged in increasing order. It can be observed that the boxes in the box plot move downward as the current increases, indicating that the error of the FDM implicit Euler method decreases as the current increases.

Figure \ref{fig:3}(b) shows the box plot distribution of MAE for the spectral method under different currents and nodes. It can be observed that the boxes are very flat, indicating that the spectral method converges quickly and its error does not change much with an increasing number of nodes. This method shows a trend similar to that of the FDM implicit Euler method, with errors decreasing as the current increases. The outlier points correspond to the spectral method with two nodes. This indicates that before the spectral method converges, its error increases with increasing current, but after convergence, the error decreases as current increases.

Figure \ref{fig:3}(c) shows the box plot distribution of MAE for the Padé approximation method under different currents and nodes. It can be seen that the error of the Padé approximation method increases significantly with increasing current. This indicates that while the Padé approximation method maintains good accuracy under low current conditions, its error increases as the current magnitude increases. This trend is opposite to that of the FDM implicit Euler method and the spectral method. Additionally, since this study uses currents up to a maximum of 2C, the Padé approximation method shows a relatively high average error.

\subsubsection{Dynamic Current Conditions}

\begin{figure}[H]
    \noindent
    \makebox[\linewidth]{\includegraphics[width=0.9\textwidth]{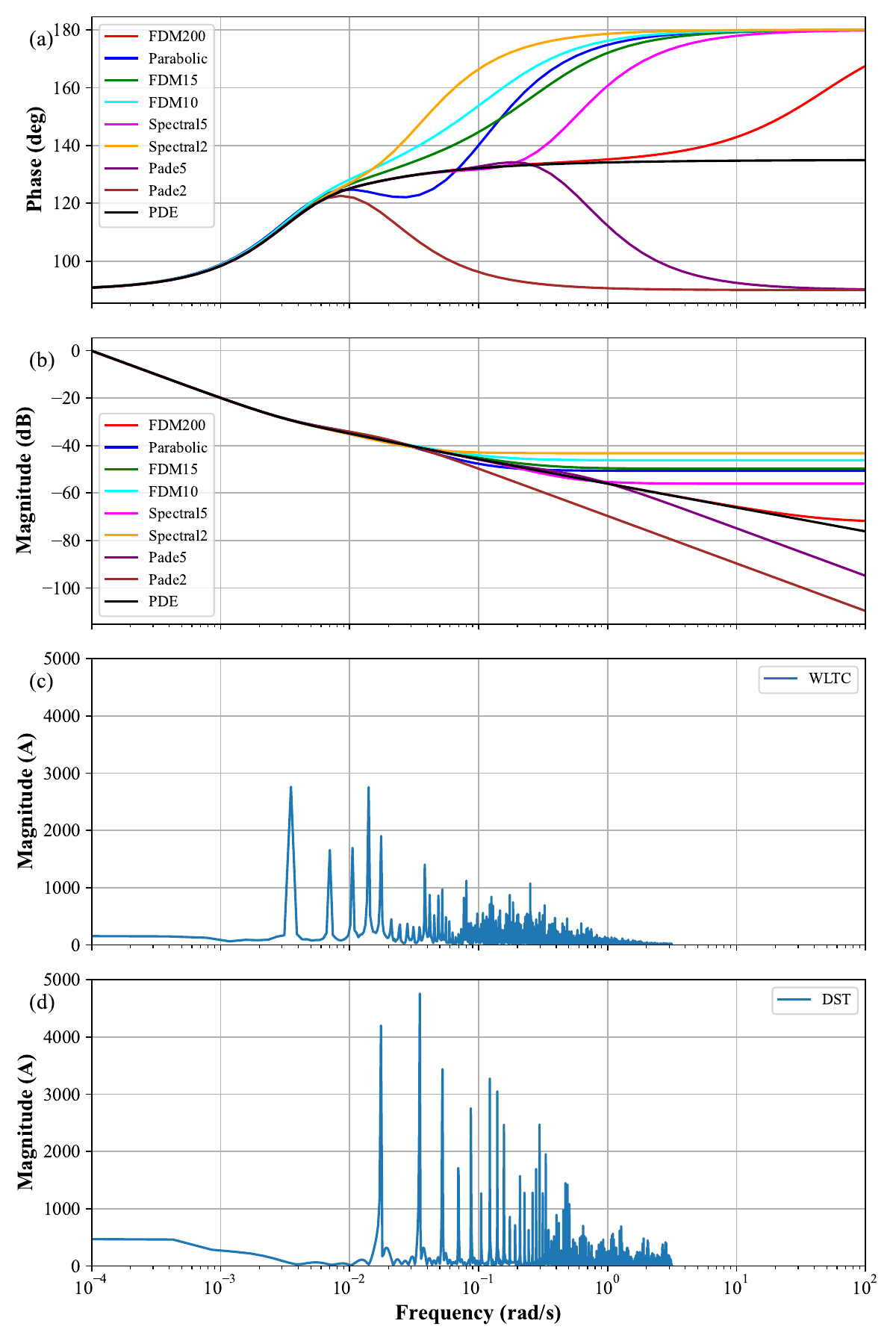}}
    \vspace{-0.8cm}
    \caption{Frequency domain Bode plot results for various methods and FFT results for WLTC and DST.}
    \label{fig:4}
\end{figure}

To analyze the performance of various methods under dynamic current conditions, this study employs frequency domain analysis. Figure \ref{fig:4}(a) and (b) present the Bode plots of different methods . The PDE represents the performance of Eq.\eqref{eq:1} as a trascendental transfer function and serves as a reference. Additionally, the models selected for comparison include the FDM implicit Euler method with 10, 15, and 200 nodes, the 2nd-order and 5th-order Padé approximation methods, the spectral method with 2 and 5 nodes, and the parabolic approximation method. 

In Figure\ref{fig:4}(a), the phase response represents the phase shift of the input signal at different frequencies by the system. The closer a curve is to the PDE solution, the better the accuracy of the method. The phase responses of the different methods begin to diverge significantly around the frequency of $10^{-2}$ rad/s. The phase response shows noticeable differences at the higher frequency range, indicating that the methods affect the phase of the signal differently even at these lower frequencies. This suggests that the phase characteristics of these methods are more sensitive to frequency changes and start to show distinct behaviors earlier compared to the magnitude response.

In Figure\ref{fig:4}(b), for magnitude response, the magnitude response represents the degree to which the system amplifies or attenuates the input signal at different frequencies. The closer a curve is to the PDE solution in the figure, the more accurate the method is. The magnitude responses of different methods start to show noticeable differences around the frequency of $10^{-1}$ rad/s. This indicates that, for lower frequency ranges (below $10^{-1}$ rad/s), these methods have very similar gain characteristics. The system’s attenuation or amplification of different frequency components is nearly identical across the methods at these low frequencies. In other words, these methods all have good accuracy at low frequencies. The differences become significant only when the frequency increases to around $10^{-1}$ rad/s, revealing distinct frequency response characteristics of each method.

Specifically, the Bode plots for each method show that the curves for the Padé approximation methods are consistently below the PDE reference solution, whereas most of the curves for the FDM, spectral method, and parabolic approximation method are above the PDE reference solution. This indicates that the Padé approximation methods and the other methods exhibit different characteristics under dynamic conditions. All methods, except for the parabolic approximation method, show an increase in accuracy under dynamic conditions as the number of nodes or the order increases.

This study uses the Fast Fourier Transform (FFT) method to perform Fourier decomposition on the dynamic conditions WLTC and DST, which are commonly used for testing models, to show their distribution at different frequencies. Since the results of FFT are related to the sampling time, generally, a smaller sampling time can reveal more dynamic characteristics. The sampling time selected for this study is 1 s, which is the same as that used in the model. The Fourier decomposition results for WLTC and DST are shown in Figure\ref{fig:4}(c) and (d). It can be observed that the highest frequency components of both dynamic conditions are cut off around 2 rad/s. Combined with Figure\ref{fig:4}(a) and (b), most models, except for FDM200, show significant errors compared to PDE near the frequency of 2 rad/s. This indicates that these models have larger errors in the high-frequency range. In practical applications, the accuracy of the models in the high-frequency range can be improved by increasing the number of nodes. 

\subsection{Execution Time}

This section introduces how the execution time among methods is evaluated and compared, followed by the obtained results.

\subsubsection{Execution Time Evaluation}
\label{sec:Execution Time Experimental Design}

This study compares the running time per step for each method. To calculate the time required for each step of a process, we use the formula which divides the total runtime by the number of steps. This is essential for understanding the efficiency of each step in the overall process. The formula is represented as follows:

\begin{equation}
T_{\text{step}} = \frac{T_{\text{total}}}{N}
    \label{eq:46}
\end{equation}
where:
\begin{itemize}
    \item \(T_{\text{step}}\) is the time required for each individual step,
    \item \(T_{\text{total}}\) is the total runtime of the process,
    \item \(N\) is the number of steps in the process.
\end{itemize}

Here, a 'step' refers to each sampling time, where the battery model runs once to obtain a data point. For example, the constant current discharge time at 0.5C is longer than at 1C, which results in a longer model runtime for 0.5C constant current discharge compared to 1C. To eliminate the impact of varying current input data lengths on computation time, we use the average computation time per step of the battery model. Therefore, a 1C constant current discharge condition was chosen to assess the execution time. 

\subsubsection{Execution Time Results}

\begin{figure}[H]
    \noindent
    \makebox[\linewidth]{\includegraphics[width=1.2\textwidth]{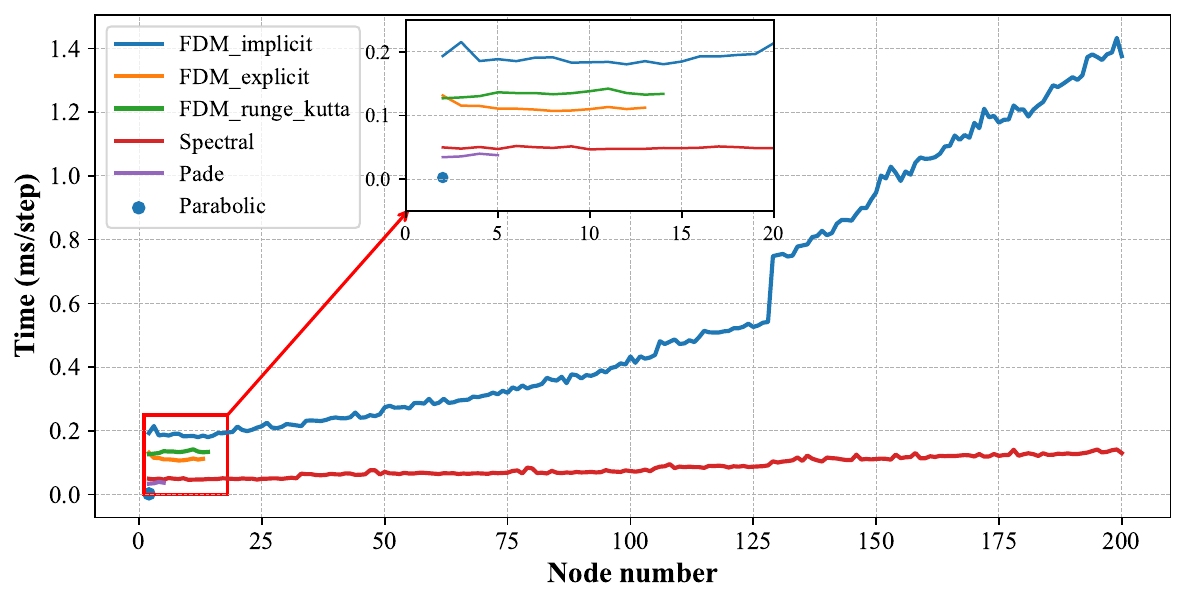}}
    \vspace{-0.8cm}
    \caption{Execution time results with different nodes.}
    \label{fig:5}
\end{figure}

The comparison results of the execution time per step for each method are shown in Figure \ref{fig:5}. The running times, sorted from longest to shortest, are: FDM implicit Euler method, FDM Runge-Kutta method, FDM explicit Euler method, spectral method, Padé approximation method, and parabolic approximation method. The specific execution time per step for each method can be found in Table \ref{tab:4}.

\begin{table}[ht]
    \centering
    \caption{Average execution time for different methods (ms/step).}
    {\footnotesize
    \begin{tabular}{ccccccc}
        \toprule
        Node & Implicit  & Explicit & Runge-Kutta & Spectral   & Padé  & Parabolic \\
        \midrule
        2  & 0.19 & 0.13 & 0.13 & 0.050 & 0.034 & 0.0025 \\
        3  & 0.22 & 0.12 & 0.13 & 0.048 & 0.036 & \\
        4  & 0.19 & 0.12 & 0.13 & 0.051 & 0.040 & \\
        5  & 0.19 & 0.11 & 0.14 & 0.047 & 0.038 & \\
        6  & 0.19 & 0.11 & 0.14 & 0.052 & & \\
        7  & 0.19 & 0.11 & 0.14 & 0.050 & & \\
        8  & 0.19 & 0.11 & 0.13 & 0.049 & & \\
        9  & 0.18 & 0.11 & 0.14 & 0.051 & & \\
        10 & 0.18 & 0.11 & 0.14 & 0.047 & & \\
        11 & 0.18 & 0.11 & 0.14 & 0.047 & & \\
        12 & 0.18 & 0.11 & 0.14 & 0.048 & & \\
        13 & 0.19 & 0.11 & 0.13 & 0.048 & & \\
        14 & 0.18 & & 0.13 & 0.049 & & \\
        15 & 0.18 & & & 0.049 & & \\
        16 & 0.19 & & & 0.049 & & \\
        17 & 0.19 & & & 0.051 & & \\
        18 & 0.20 & & & 0.050 & & \\
        19 & 0.20 & & & 0.049 & & \\
        20 & 0.21 & & & 0.049 & & \\
        \bottomrule
    \end{tabular}
    }
    \label{tab:4}
\end{table}

The results show that the parabolic approximation method has the fastest running speed, being 10 times faster than the second fastest, the Padé approximation method. This indicates that if very fast running speeds are required, such as for application in BMS or aging simulations, the parabolic approximation method is the optimal choice. The FDM implicit Euler method has the slowest running speed, which increases significantly with the number of nodes. This is because the implicit Euler method involves matrix inversion, which has a time complexity of \( O(n^3) \). In contrast, the spectral method requires less computation time than the FDM methods, and its computation time grows very slowly with the number of nodes. Moreover, since the spectral method with 5 nodes already achieves high accuracy, it offers faster computation speeds compared to the FDM methods when the same level of accuracy is required.

\subsection{Memory Usage}
This section shows first how the memory usage is evaluated and compared for the different considered methods, followed by the obtained results.

\subsubsection{Memory Usage Evaluation}

 To measure the memory usage of model execution, we employed the \texttt{tracemalloc} module in Python, using the peak memory usage as the metric. The memory usage calculation includes all variables and data required for model execution. Since the selection and storage method of variables directly impact memory usage, we maintained a consistent programming style and stored identical content in the same manner. Additionally, when using electrochemical models to obtain critical information about the battery' internal behaviour, we recorded the output voltage, the average lithium-ion concentration, and the surface lithium-ion concentration in both the positive and negative electrodes for each method. These values were stored in arrays for output. A 1C constant current discharge condition was chosen to assess the memory usage. To minimize the randomness in memory usage for each run, we ran each model 10 times and used the average memory consumption from these runs. This approach helps smooth out the variability in memory usage and provides a more reliable measure of the typical memory requirements for each method.

\subsubsection{Memory Usage Results}

\begin{figure}[H]
    \noindent
    \makebox[\linewidth]{\includegraphics[width=1.2\textwidth]{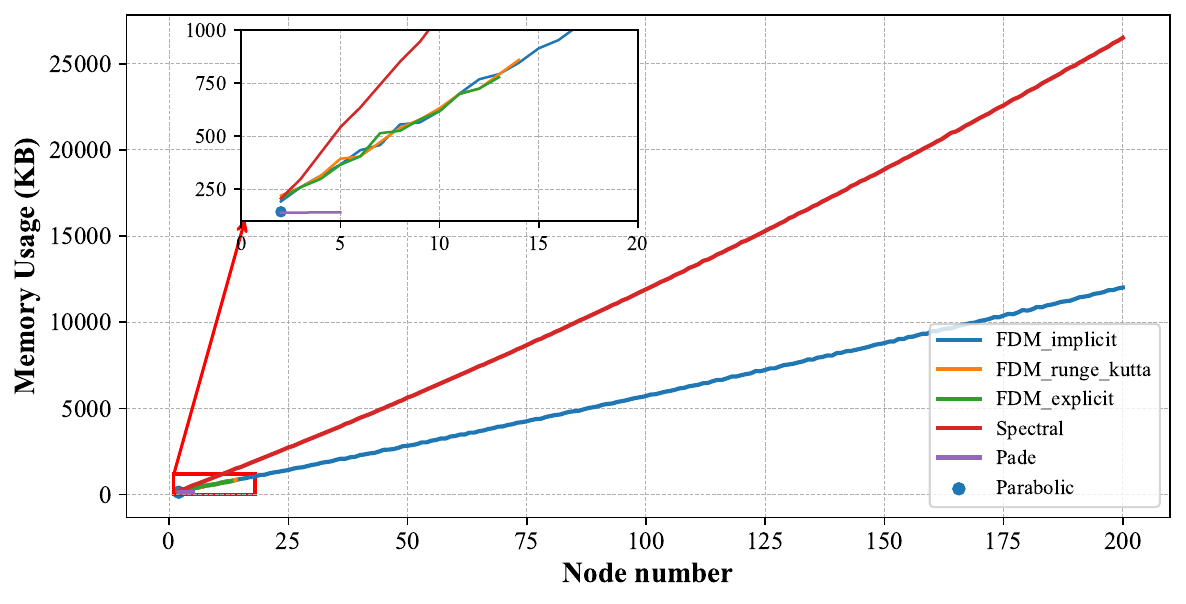}}
    \vspace{-0.8cm}
    \caption{Memory usage results with different nodes.} 
    \label{fig:6}
\end{figure}

From Figure \ref{fig:6} and Table \ref{tab:5}. These results show that the spectral method has the highest memory consumption. This is due to the use of Chebyshev polynomials in approximating the PDE solutions, which introduces more variables that need to be stored. However, the accuracy results indicate that even with only 5 nodes, the spectral methods achieve good accuracy and computational efficiency. Therefore, in practical applications, it is unnecessary to use a large number of nodes, such as 200, which helps keeping memory usage within a reasonable range. For instance, the memory usage of the spectral method with 5 nodes is even lower than that of the FDM implicit Euler method with 10 nodes, yet it offers better accuracy. The memory usage of the three FDM methods is almost the same, suggesting that different solution techniques do not significantly affect memory consumption. The memory usage of the Padé approximation method has the lowest memory consumption and remains nearly constant, indicating that increasing the order has little impact on its memory consumption. The parabolic approximation method shows similar memory usage to the Padé approximation method.

\begin{table}[H]
    \centering
    \caption{Memory usage for different methods (KB).}
    \label{tab:5}
{\footnotesize
    \begin{tabular}{ccccccc}
        \toprule
        Node & Implicit & Explicit & Runge-Kutta & Spectral & Padé & Parabolic \\
        \midrule
         2  & 192.54 & 206.06 & 220.17 & 206.69 & 139.93 & 143.48 \\
         3  & 260.00 & 259.14 & 259.16 & 298.77 & 140.08 & \\
         4  & 313.12 & 298.18 & 312.28 & 420.55 & 140.26 & \\
         5  & 366.26 & 365.40 & 393.61 & 542.10 & 140.47 & \\
         6  & 433.90 & 404.51 & 404.51 & 634.99 & & \\
         7  & 458.20 & 513.98 & 471.82 & 742.61 & & \\
         8  & 554.81 & 525.02 & 539.17 & 850.48 & & \\
         9  & 564.71 & 578.29 & 578.34 & 944.17 & & \\
        10  & 618.02 & 617.54 & 631.65 & 1067.04 & & \\
        11  & 700.27 & 699.03 & 699.11 & 1175.72 & & \\
        12  & 768.09 & 724.24 & 724.24 & 1270.21 & & \\
        13  & 792.58 & 777.64 & 791.77 & 1393.88 & & \\
        14  & 846.01 & & 859.35 & 1517.77 & & \\
        15  & 913.92 & & & 1598.64 & & \\
        16  & 952.96 & & & 1723.10 & & \\
        17  & 1020.94 & & & 1833.38 & & \\
        18  & 1045.58 & & & 1943.92 & & \\
        19  & 1142.55 & & & 2054.72 & & \\
        20  & 1153.38 & & & 2151.33 & & \\
        \bottomrule
    \end{tabular}
}
\end{table}

\subsection{Overall Performance Evaluation}

\begin{figure}[H]
    \noindent
    \makebox[\linewidth]{\includegraphics[width=1.2\textwidth]{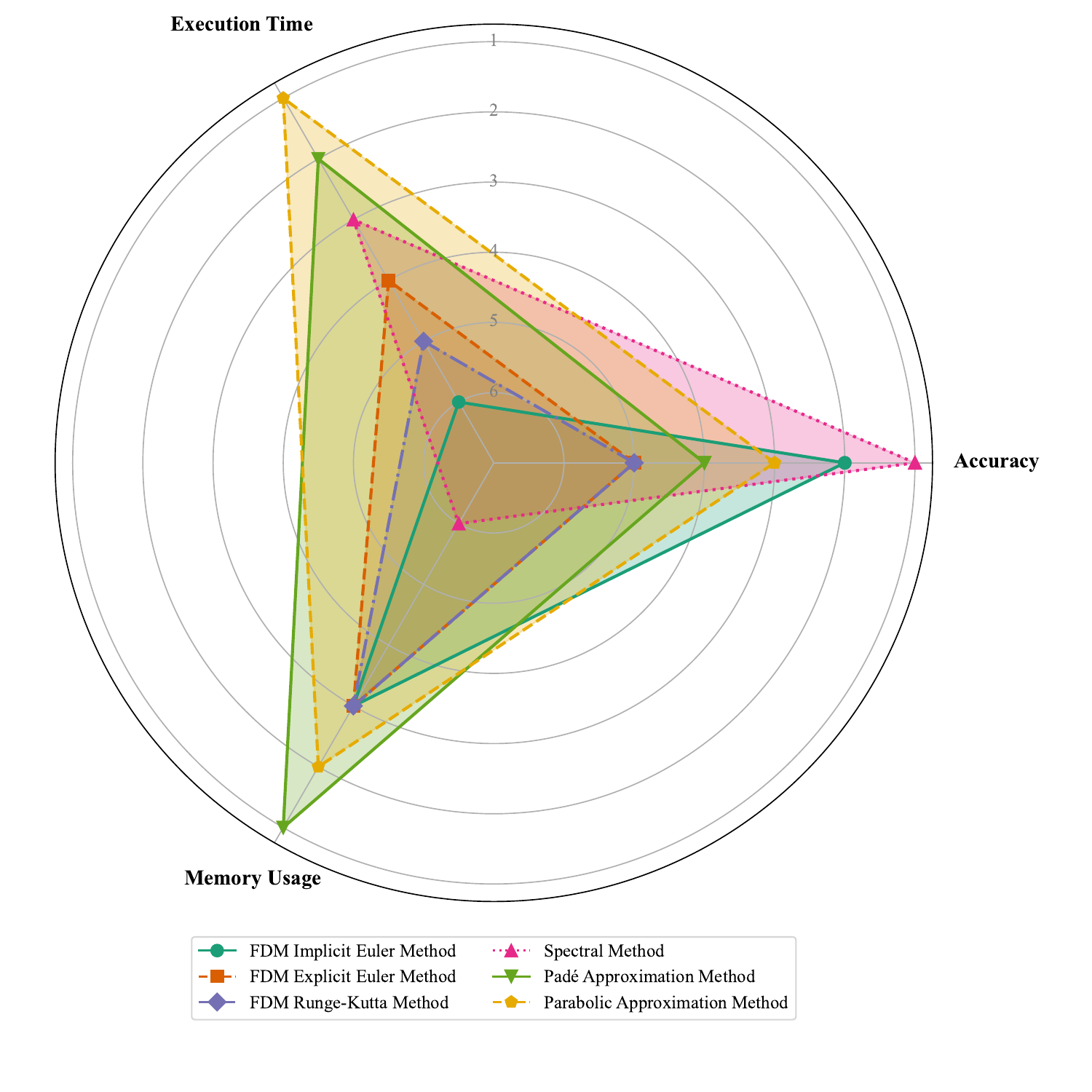}}
    \vspace{-0.8cm}
    \caption{Model performance comparison.}
    \label{fig:7}
\end{figure}

Figure \ref{fig:7} shows the rankings of various methods in terms of accuracy, computation time, and memory usage. It shows that the FDM explicit Euler method and the FDM Runge-Kutta method do not perform well in any of these three dimensions and can be avoided in practical applications. The benchmark FDM implicit Euler method achieves good accuracy with an increased number of nodes, but this also results in higher computation time and memory usage. The spectral method offers the best accuracy and convergence speed, with computation time also better than the FDM implicit method, though its memory usage is relatively high.

The Padé approximation method has the lowest memory usage and ranks second in computation time, but its accuracy is lower compared to the spectral method, FDM implicit method, and parabolic method. Additionally, the Padé approximation method shows increased errors under high constant current conditions, and increasing its order does not improve accuracy under these conditions. However, for dynamic conditions, increasing the order can better capture the high-order response of the model.

The parabolic approximation method demonstrates the best computation time, being 10 times faster than the second-best Padé approximation method. Its memory usage is also relatively low, and it provides better accuracy under constant current conditions than the Padé approximation method. However, its drawback is that its accuracy cannot be further improved by increasing the number of nodes, as can be done with other methods. 

In practical applications, the choice of method can be based on the rankings in these three dimensions. For example, in digital twin applications, where high-fidelity offline simulations are often required to support accurate diagnostics, prediction, and system optimization, model accuracy is the primary concern, while computational speed and memory consumption are less critical due to the availability of sufficient computing resources. So, the spectral method or FDM implicit Euler method can be use. If model accuracy is crucial and faster execution is desired, such as in battery aging simulations, the spectral method can be used. For simulations that consider extreme conditions or need to include more dynamic characteristics of the model, increasing the number of nodes can enhance accuracy and dynamic response. For applications like BMS or other embedded system  that require faster computation speed and lower memory usage, the parabolic approximation method or the Padé approximation method can be employed. However, if the model's conditions include high current scenarios, it is advisable to avoid using the Padé approximation method.

\section{Conclusion}
\label{sec:Conclu}
This study compares several commonly used numerical discretization methods in solving the SPM model, including FDM (implicit Euler method, explicit Euler method, and Runge-Kutta method), spectral methods, Padé approximation, and parabolic approximation. The evaluation focuses on accuracy, execution time, and memory usage. It aims to provide guidance on the selection of solving methods for electrochemical models.

Under constant current conditions, the FDM explicit Euler method and Runge-Kutta method exhibit relatively large errors. The FDM implicit Euler method offers higher accuracy compared to these two methods and can continuously improve accuracy with an increasing number of nodes. The spectral method provides the best accuracy and converges faster  than the FDM implicit method, achieving good results with just five nodes. Both the FDM implicit method and the spectral method show decreasing errors with increasing current magnitudes. The Padé approximation method has larger errors compared to the FDM implicit Euler method, spectral method, and parabolic approximation method. Additionally, its error increases significantly with higher currents. The parabolic method's error is greater than that of the converged spectral method and the FDM implicit Euler method. Our frequency domain analysis of various models and dynamic conditions indicates that the FDM method, spectral method, and Padé approximation method can improve high-frequency response accuracy by increasing the number of nodes or the order of the method.

Comparison of the execution time results from different methods shows that the parabolic approximation method has the fastest execution speed, followed by the Pade approximation method. The spectral method is faster than the FDM methods, with the FDM implicit Euler method being the slowest. Regarding memory usage, the parabolic approximation method and Pade approximation method use the least memory, while the various FDM solving methods have minimal memory usage. The spectral method has the highest memory usage. Future researchers can use the results of this study, along with specific application scenarios, to select appropriate numerical discretization methods for solving electrochemical models.

\section*{Acknowledgements}

The authors declare that they have no acknowledgements to report.
% \section*{Acknowledgements}
% We would like to thank [Name or Institution] for [specific help or funding].

\section*{Conflict of Interest}
The authors declare that they have no conflict of interest.

%% The Appendices part is started with the command \appendix;
%% appendix sections are then done as normal sections

%% If you have bibdatabase file and want bibtex to generate the
%% bibitems, please use
%%
 \bibliographystyle{elsarticle-num} 
 \bibliography{cas-refs}

\begin{thebibliography}{10}
\expandafter\ifx\csname url\endcsname\relax
  \def\url#1{\texttt{#1}}\fi
\expandafter\ifx\csname urlprefix\endcsname\relax\def\urlprefix{URL }\fi
\expandafter\ifx\csname href\endcsname\relax
  \def\href#1#2{#2} \def\path#1{#1}\fi

\bibitem{jiang2024advances}
M.~Jiang, D.~Li, Z.~Li, Z.~Chen, Q.~Yan, F.~Lin, C.~Yu, B.~Jiang, X.~Wei, W.~Yan, et~al., Advances in battery state estimation of battery management system in electric vehicles, Journal of Power Sources 612 (2024) 234781.

\bibitem{waseem2023battery}
M.~Waseem, M.~Ahmad, A.~Parveen, M.~Suhaib, Battery technologies and functionality of battery management system for evs: Current status, key challenges, and future prospectives, Journal of Power Sources 580 (2023) 233349.

\bibitem{guo2024systematic}
F.~Guo, L.~D. Couto, G.~Mulder, K.~Trad, G.~Hu, O.~Capron, K.~Haghverdi, A systematic review of electrochemical model-based lithium-ion battery state estimation in battery management systems, Journal of Energy Storage 101 (2024) 113850.

\bibitem{li2021model}
Y.~Li, D.~Karunathilake, D.~M. Vilathgamuwa, Y.~Mishra, T.~W. Farrell, C.~Zou, et~al., Model order reduction techniques for physics-based lithium-ion battery management: A survey, IEEE Industrial Electronics Magazine 16~(3) (2021) 36--51.

\bibitem{ali2024comparison}
H.~A.~A. Ali, L.~H. Raijmakers, K.~Chayambuka, D.~L. Danilov, P.~H. Notten, R.-A. Eichel, A comparison between physics-based li-ion battery models, Electrochimica Acta (2024) 144360.

\bibitem{fuller1994simulation}
T.~F. Fuller, M.~Doyle, J.~Newman, Simulation and optimization of the dual lithium ion insertion cell, Journal of the electrochemical society 141~(1) (1994) 1.

\bibitem{haran1998determination}
B.~S. Haran, B.~N. Popov, R.~E. White, Determination of the hydrogen diffusion coefficient in metal hydrides by impedance spectroscopy, Journal of Power Sources 75~(1) (1998) 56--63.

\bibitem{marquis2019asymptotic}
S.~G. Marquis, V.~Sulzer, R.~Timms, C.~P. Please, S.~J. Chapman, An asymptotic derivation of a single particle model with electrolyte, Journal of The Electrochemical Society 166~(15) (2019) A3693.

\bibitem{di2010lithium}
D.~Di~Domenico, A.~Stefanopoulou, G.~Fiengo, Lithium-ion battery state of charge and critical surface charge estimation using an electrochemical model-based extended kalman filter, Journal of Dynamic Systems, Measurement, and Control 132 (2010) 061302--1.

\bibitem{lotfi2020switched}
F.~Lotfi, S.~Ziapour, F.~Faraji, H.~D. Taghirad, A switched sdre filter for state of charge estimation of lithium-ion batteries, International Journal of Electrical Power \& Energy Systems 117 (2020) 105666.

\bibitem{xiong2018electrochemical}
R.~Xiong, L.~Li, Z.~Li, Q.~Yu, H.~Mu, An electrochemical model based degradation state identification method of lithium-ion battery for all-climate electric vehicles application, Applied energy 219 (2018) 264--275.

\bibitem{plante2022multiple}
E.~Plant{\'e}, R.~Postoyan, S.~Ra{\"e}l, Y.~Jebroun, S.~Benjamin, D.~M. Reyes, Multiple active material lithium-ion batteries: Finite-dimensional modeling and constrained state estimation, IEEE Transactions on Control Systems Technology 31~(3) (2022) 1106--1121.

\bibitem{wang2023lithium}
J.~Wang, J.~Meng, Q.~Peng, T.~Liu, X.~Zeng, G.~Chen, Y.~Li, Lithium-ion battery state-of-charge estimation using electrochemical model with sensitive parameters adjustment, Batteries 9~(3) (2023) 180.

\bibitem{sturm2018state}
J.~Sturm, H.~Ennifar, S.~V. Erhard, A.~Rheinfeld, S.~Kosch, A.~Jossen, State estimation of lithium-ion cells using a physicochemical model based extended kalman filter, Applied energy 223 (2018) 103--123.

\bibitem{li2023physics}
Y.~Li, Z.~Wei, C.~Xie, D.~M. Vilathgamuwa, Physics-based model predictive control for power capability estimation of lithium-ion batteries, IEEE Transactions on Industrial Informatics (2023).

\bibitem{tian2023aging}
A.~Tian, C.~Yang, Y.~Gao, Y.~Jiang, C.~Chang, L.~Wang, J.~Jiang, Aging effect--aware finite element model and parameter identification method of lithium-ion battery, Journal of Electrochemical Energy Conversion and Storage 20~(3) (2023) 031005.

\bibitem{yeregui2023state}
J.~Yeregui, L.~Oca, I.~Lopetegi, E.~Garayalde, M.~Aizpurua, U.~Iraola, State of charge estimation combining physics-based and artificial intelligence models for lithium-ion batteries, Journal of Energy Storage 73 (2023) 108883.

\bibitem{trefethen2000spectral}
L.~N. Trefethen, Spectral methods in MATLAB, SIAM, 2000.
\newblock \href {https://doi.org/https://epubs.siam.org/doi/book/10.1137/1.9780898719598} {\path{doi:https://epubs.siam.org/doi/book/10.1137/1.9780898719598}}.

\bibitem{bizeray2015lithium}
A.~M. Bizeray, S.~Zhao, S.~R. Duncan, D.~A. Howey, Lithium-ion battery thermal-electrochemical model-based state estimation using orthogonal collocation and a modified extended kalman filter, Journal of Power Sources 296 (2015) 400--412.

\bibitem{wang2023system}
Y.~Wang, X.~Zhang, K.~Liu, Z.~Wei, X.~Hu, X.~Tang, Z.~Chen, System identification and state estimation of a reduced-order electrochemical model for lithium-ion batteries, eTransportation 18 (2023) 100295.

\bibitem{forman2010reduction}
J.~C. Forman, S.~Bashash, J.~L. Stein, H.~K. Fathy, Reduction of an electrochemistry-based li-ion battery model via quasi-linearization and pade approximation, Journal of the Electrochemical Society 158~(2) (2010) A93.

\bibitem{hosseininasab2023state}
S.~Hosseininasab, N.~Momtaheni, S.~Pischinger, M.~G{\"u}nther, L.~Bauer, State-of-charge estimation of lithium-ion batteries using an adaptive dual unscented kalman filter based on a reduced-order model, Journal of Energy Storage 73 (2023) 109011.

\bibitem{fang2023performance}
D.~Fang, W.~Wu, J.~Li, W.~Yuan, T.~Liu, C.~Dai, Z.~Wang, M.~Zhao, Performance simulation method and state of health estimation for lithium-ion batteries based on aging-effect coupling model, Green Energy and Intelligent Transportation 2~(3) (2023) 100082.

\bibitem{subramanian2005efficient}
V.~R. Subramanian, V.~D. Diwakar, D.~Tapriyal, Efficient macro-micro scale coupled modeling of batteries, Journal of The Electrochemical Society 152~(10) (2005) A2002.

\bibitem{romero2011comparison}
A.~Romero-Becerril, L.~Alvarez-Icaza, Comparison of discretization methods applied to the single-particle model of lithium-ion batteries, Journal of Power Sources 196~(23) (2011) 10267--10279.

\bibitem{xu2023comparative}
L.~Xu, J.~Cooper, A.~Allam, S.~Onori, Comparative analysis of numerical methods for lithium-ion battery electrochemical modeling, Journal of The Electrochemical Society 170~(12) (2023) 120525.

\bibitem{bizeray_2016_212178}
A.~M. Bizeray, J.~Reniers, D.~A. Howey, \href{https://doi.org/10.5281/zenodo.212178}{Spectral\_li-ion\_spm: Initial release} (Dec. 2016).
\newblock \href {https://doi.org/10.5281/zenodo.212178} {\path{doi:10.5281/zenodo.212178}}.
\newline\urlprefix\url{https://doi.org/10.5281/zenodo.212178}

\bibitem{pechlivanidou2022zero}
G.~Pechlivanidou, N.~Karampetakis, Zero-order hold discretization of general state space systems with input delay, IMA Journal of Mathematical Control and Information 39~(2) (2022) 708--730.

\bibitem{marcicki2013design}
J.~Marcicki, M.~Canova, A.~T. Conlisk, G.~Rizzoni, Design and parametrization analysis of a reduced-order electrochemical model of graphite/lifepo4 cells for soc/soh estimation, Journal of Power Sources 237 (2013) 310--324.

\bibitem{moura2014adaptive}
S.~J. Moura, N.~A. Chaturvedi, M.~Krsti{\'c}, Adaptive partial differential equation observer for battery state-of-charge/state-of-health estimation via an electrochemical model, Journal of Dynamic Systems, Measurement, and Control 136~(1) (2014) 011015.

\bibitem{zhang2022beyond}
D.~Zhang, S.~Park, L.~D. Couto, V.~Viswanathan, S.~J. Moura, Beyond battery state of charge estimation: Observer for electrode-level state and cyclable lithium with electrolyte dynamics, IEEE Transactions on Transportation Electrification (2022).

\bibitem{li2022unlocking}
W.~Li, Y.~Fan, F.~Ringbeck, D.~J{\"o}st, D.~U. Sauer, Unlocking electrochemical model-based online power prediction for lithium-ion batteries via gaussian process regression, Applied Energy 306 (2022) 118114.

\bibitem{tagade2016recursive}
P.~Tagade, K.~S. Hariharan, P.~Gambhire, S.~M. Kolake, T.~Song, D.~Oh, T.~Yeo, S.~Doo, Recursive bayesian filtering framework for lithium-ion cell state estimation, Journal of Power Sources 306 (2016) 274--288.

\bibitem{fan2020systematic}
G.~Fan, Systematic parameter identification of a control-oriented electrochemical battery model and its application for state of charge estimation at various operating conditions, Journal of Power Sources 470 (2020) 228153.

\bibitem{guo2024efficiency}
F.~Guo, L.~D. Couto, G.~Thenaisie, Efficiency and optimality in electrochemical battery model parameter identification: A comparative study of estimation techniques, in: 2024 10th International Conference on Optimization and Applications (ICOA), IEEE, 2024, pp. 1--6.

\end{thebibliography}

%% else use the following coding to input the bibitems directly in the
%% TeX file.

% \begin{thebibliography}{00}

% %% \bibitem{label}
% %% Text of bibliographic item

% \bibitem{}

% \end{thebibliography}
\end{document}